\documentclass[preprint,aps]{revtex4}

\usepackage{graphicx}
\usepackage{dcolumn}
\usepackage{bm}

\usepackage{epsf}
\usepackage{anysize}
\usepackage{graphicx}
\marginsize{1 in}{1 in}{1in}{1.40in}

\usepackage{graphicx}
\usepackage{dcolumn}
\usepackage{bm}
\usepackage{amsmath}
\usepackage{amssymb}
\usepackage{mathrsfs}
\usepackage[latin1]{inputenc}
\usepackage{amsmath}

\makeatletter
  \newcommand\figcaption{\def\@captype{figure}\caption}
  \newcommand\tabcaption{\def\@captype{table}\caption}
\makeatother

\newcommand{\dd}{{\rm d\,}}
\newcommand{\e}{\epsilon}

\newcommand{\sd}{Schr\"{o}dinger }

\newcommand{\lmd}{\lambda }
\newcommand{\tr}{{\rm Tr}}

\newcommand{\K}{\mathbb{K}}
\newcommand{\U}{\mathcal{U}}

\newcommand{\M}{\mathcal{M}}
\newcommand{\h}{\mathcal{H}}
\newcommand{\X}{\mathcal{X}}
\newcommand{\Y}{\mathcal{Y}}

\newcommand{\q}{{\varrho}}
\newcommand{\F}{\mathscr{F}}

\newtheorem{theorem}{Theorem}

\begin{document}

\title{Control Landscapes for Observable Preparation with Open Quantum Systems}

\author{Rebing Wu}
\altaffiliation{Department of Chemistry, Princeton University,
Princeton, New Jersey 08544, USA}


\author{Alexander Pechen}
\altaffiliation{Department of Chemistry, Princeton University,
Princeton, New Jersey 08544, USA}

\author{Herschel Rabitz}
\altaffiliation{Department of Chemistry, Princeton University,
Princeton, New Jersey 08544, USA}

\author{Michael Hsieh}
\altaffiliation{Department of Chemistry, Princeton University,
Princeton, New Jersey 08544, USA}

\author{Benjamin Tsou}
\altaffiliation{Department of Chemistry, Princeton University,
Princeton, New Jersey 08544, USA}


\begin{abstract}
A quantum control landscape is defined as the observable as a
function(al) of the system control variables. Such landscapes were
introduced to provide a basis to understand the increasing number
of successful experiments controlling quantum dynamics phenomena.
This paper extends the concept to encompass the broader context of
the environment having an influence. For the case that the open
system dynamics are fully controllable, it is shown that the
control landscape for open systems can be lifted to the analysis
of an equivalent auxiliary landscape of a closed composite system
that contains the environmental interactions. This inherent
connection can be analyzed to provide relevant information about
the topology of the original open system landscape. Application to
the optimization of an observable expectation value reveals the
same landscape simplicity observed in former studies on closed
systems. In particular, no false sub-optimal traps exist in the
system control landscape when seeking to optimize an observable,
even in the presence of complex environments. Moreover, a
quantitative study of the control landscape of a system
interacting with a thermal environment shows that the enhanced
controllability attainable with open dynamics significantly
broadens the range of the achievable observable values over the
control landscape.
\end{abstract}

\maketitle
\section{Introduction}
Quantum optimal control concepts are being used to address many
applications including bond-selective chemical reactions, quantum
state engineering, and quantum computation
\cite{rabitz1,Rabitz2000,weitz,dantus,vlasta,lloyd2}. Toward the
natural goal of finding the best attainable outcome, optimal
control theory can be applied to redirect the quantum dynamics
subject to any practical constraints on the laboratory controls
\cite{Wusheng1998}. In the laboratory, optimal control experiments
generally have been based on learning algorithms \cite{rabitz9}
that guide a closed-loop search for an effective control despite
the lack of full information about the system's Hamiltonian, its
interaction with the environment and the influence of noise. The
growing number of experimental successes
\cite{meshulach,levis,bixner} collectively show that it is
(unexpectedly) easy to find high-quality control results even with
severely constrained control fields (i.e., shaped pulses from the
same Ti:sappire laser operating at $\sim$800nm with a bandwidth of
$\sim$20nm are used as the driver for most of the experiments).

The quantum control landscape is the expectation value of the
physical observable as a functional $J = J[\e]$ of the time
dependent control field $\e(\cdot)$. Optimal controls maximize (or
minimize, depending on the control goal) the objective functional,
i.e., such that $J[\e] = \max_\e J[\e]$ (resp., $J[\e] = \min_\e
J[\e]$). In the laboratory, for controlling a complex system
typically detailed knowledge about the dependence of the objective
functional on the control is not available, and the practical
search for an optimal control solution is performed adaptively via
various algorithms (e.g., genetic algorithms, gradient methods).
If the objective functional has local traps (local minima or
maxima), then their presence can slow down the efficiency of the
search or even permanently trap the algorithm in a local maximum
(resp., minimum). This circumstance motivates the analysis of the
topological properties of quantum control landscapes. In
particular, {\it a priori} information about the absence of local
traps for common quantum objective functionals for a wide class of
quantum systems would imply that the laboratory searches would not
be hindered and a global optimal solution would be finally reached
(assuming that the control field is sufficiently flexible).
Moreover, the lack of landscape traps would provide an explanation
for the evident relative practical ease of obtaining optimal
solutions. Recent studies
\cite{rabitz6,rabitz10,rabitz11,tak-san,mike,dmorph,mike2,rebing}
show that such favorable landscape properties exist for optimizing
the expectation value of a system observable for any controllale
finite level closed quantum system.

Realistic quantum systems are always exposed to some kind of
environment (e.g., chemical reactions in heat baths, or atoms in
optical cavities), and their landscape analysis for such systems
is important. In most cases, the environmentally induced
irreversible quantum dynamics \cite{alicki,chuang} would downgrade
the quality of the control outcome, especially when little can be
done to tailor the environment. This situation can be serious in
many applications, especially quantum information sciences
\cite{chuang}. However, there are indications that it may be
possible to fight against or even cooperate with environmentally
induced dynamics \cite{shuang}. Moreover, combining incoherent
control by the environment with coherent control by an
electromagnetic field offers a general technique for manipulating
quantum systems by optimally affecting both the Hamiltonian and
dissipative aspects of the dynamics; the latter circumstance
corresponds to the environment consisting of incoherent radiation
or a gas of electrons, atoms or molecules whose state is suitably
optimized~\cite{pechen} (e.g., to allow for the creation of mixed
target states). There also have been explicit applications of
controlled quantum correlations induced to non-classically
manipulate quantum states (e.g., through quantum error correction
using redundant qubits \cite{qec1,qec2}, atomic control in
single-mode optical cavities \cite{cavity}, and real-time feedback
control \cite{mabuchi1,mabuchi2}). These scenarios show the
prospects of attaining full control of the system evolution by
``freezing" or possibly counter-intuitively ``reversing" what was
thought to be irreversible dynamics \cite{lloyd2,brockett1}. In
the laboratory, even if an explicit treatment of the environment
is not considered, under optimization of the observable the
control will naturally address the environment to maximally draw
on its beneficial features or minimize its deleterious effects to
best achieve control of the system.

This perspective forms the background for the present paper aiming
to investigate control landscapes of open quantum systems whose
dynamics can be effectively manipulated. The landscape analyses
will be explored within the framework of the Kraus operator-sum
representation of open system dynamics \cite{pechen1}, revealing
that no local suboptima exist in the control landscape of open
systems, although there is an increase of the number of saddle
submanifolds from that in the case of analogous closed quantum
systems. The open system landscape analysis indicates that the
search for optimal controls in the laboratory may not be
significantly hindered by the presence of an environment, thereby
providing a basis to understand the relative ease of obtaining
optimal control of open systems and allowing for the development
of more effective searching algorithms.

The paper is organized as follows. Section II summarizes the
properties of Kraus maps, while section III presents the concept
of landscape lifting and discusses its general aspects. Section IV
applies landscape lifting to Kraus maps for manipulation of an
observable, thereby revealing a rich structure for the critical
submanifolds in the search space and the quantitative influence of
the environment on the landscape. Section V summarizes the
results.

\section{Representations of open quantum system dynamics}
We summarize the Kraus operator-sum formalism of open quantum
system dynamics via a simple ``system plus environment" model,
where the quantum system with Hilbert space dimension $N$ is
coupled to an external quantum environment (taken as
$\lambda$-dimensional), and the composite of the system and the
environment obeys the \sd equation,
\begin{equation}\label{total}
i\hbar\frac{\dd \rho_{\rm total}}{\dd t}=[H_{\rm total},\rho_{\rm
total}],
\end{equation}
where $\rho_{\rm total}$ is the total density matrix. The total
Hamiltonian $H_{\rm total}$ includes the internal Hamiltonians of
the system and the environment as well as their interaction, all
contributing to the total system evolution operator $U_{\rm
total}(t)$ on the total Hilbert space $\h=\h_S\otimes\h_E$, where
$\h_S$ and $\h_E$ are the Hilbert spaces of the system and
environment, respectively. The composite system is assumed to
initially be prepared in a product state $\rho_{\rm
total}(0)=\rho\otimes \q$, where $\rho$ and $\q$ are the initial
states of the system and the environment, respectively. The system
dynamics described by $\rho_S(t)$ can be obtained by tracing
$\rho_{\rm total}(t)=U_{\rm total}(t)\rho_{\rm
total}(0)U^\dagger_{\rm total}(t)$ over the environment, leading
to the Kraus operator-sum representation:
\begin{equation}\label{reduce}
\rho_S(t)=\sum_{\alpha,\beta=1}^\lambda K_{\alpha\beta}(t) \rho
K_{\alpha\beta}^\dagger(t),
\end{equation}
where $K_{\alpha\beta}(t)=\tr_E\left\{U_{\rm total}(t)(I_N\otimes
\q^{1/2}|\beta\rangle\langle \alpha|)\right\}$ is the
$(\alpha,\beta)$-th $N\times N$ block of the $\lmd N\times \lmd N$
matrix $K=U_{\rm total}(t)(I_N\otimes \q^{1/2})$ under some
arbitrary orthonormal basis
$\{|\alpha\rangle,|\beta\rangle,\cdots\}$ of $\h_E$. The matrices
$\{K_{\alpha\beta}\}$ form the Kraus representation of the
dynamical map (\ref{reduce}). For compactness, we use $K$ to
denote the Kraus representation by the set of Kraus operators
$\{K_{\alpha\beta}\}$ associated with the initial environmental
state $\q$, which always satisfies the identity
\begin{equation}\label{kraus}
K^\dagger K=I_N\otimes \q
\end{equation}
to guarantee that the trace-preservation of the system density
matrix \cite{kraus}. The set of Kraus representations induced by
the environment state $\q$ is denoted by $\K[\q,N]$, which can be
taken as the image of the unitary group $\U(\lmd N)$ under the
mapping
\begin{equation}\label{map}K=\F_\q(U)=U(I_N\otimes \q^{1/2}),~~~~U\in\U(\lmd N).\end{equation}
Suppose that the rank of $\q$ is $\mu$, then $\K[\q,N]$ is
homeomorphic to the following homogeneous space of $\U(\lmd N)$
\begin{equation}\label{coset}
\left\{U\left(\begin{array}{cc}
  I_{\mu N} &  \\
   & 0_{(\lmd-\mu) N} \\
\end{array}\right),\quad U\in \U(\lmd N)\right\}=\frac{\U(\lmd N)}{\U[(\lmd-\mu)N]},
\end{equation}
where each $K\in\K[\q,N]$ corresponds to a coset in the unitary
group.

The definition (\ref{kraus}) of Kraus representation is still
ambiguous because the reduction (\ref{reduce}) to the system
density matrix is dependent on the choice of basis for $\h_E$. In
general, two Kraus operators $K$ and $\tilde K$ correspond to the
same Kraus map if
$$\sum_{\alpha,\beta} {
K_{\alpha\beta} \rho
K_{\alpha\beta}^\dagger}=\sum_{\alpha',\beta'} {\tilde
K_{\alpha'\beta'} \rho \tilde K_{\alpha'\beta'}^\dagger},\quad
\forall \,\,\rho.$$It is easy to show that the following set of
Kraus representations
\begin{equation}\label{equivalence} \big\{\tilde K=(I_N\otimes
V)K,\quad V\in \U(\lmd)\big\}
\end{equation}
are equivalent to $K$, and they exhaust all equivalent
representations when $\q$ is a pure state. Here the unitary matrix
$V$ represents all possible basis transformations on the
environmental subsystem.

Since the sequential order of the Kraus operators in the sum does
not affect the operation of Kraus maps on the density matrix, and
the effective number of independent Kraus operators is at most
$N^2$ because the Kraus maps are linear in the space of
$N$-dimensional Hermitian operators, any Kraus map can be
associated with an $N^2$-dimensional ``fictitious" environment
starting from a pure state. This characterizes the whole set of
open system Kraus maps without any explicit environmental details
appeared in the formulation. In more general cases, $\K[\q,N]$ has
to be associated with a specific environment initial state $\q$
and the set only involves a proper subset of all possible
dynamical maps of the underlying open quantum system.

\section{Landscape lifting}
The quantum control landscape is defined by a cost functional
$J[\varepsilon(t)]$ built on the space of time-dependent control
fields that enter into the Hamiltonian $H_{\rm total}$ to
manipulate the quantum dynamics. A main objective of the analysis
is to determine the topology of the landscape extremal controls
(i.e., critical points of the landscape function) and understand
the influence of the topology upon the search for optimal
controls.

It can be difficult to carry out a landscape analysis in the space
of time-dependent control fields that form a subset of the
infinite dimensional space $\mathcal{L}^2(\mathbb{R})$ of Lebesque
integrable functions. However, under some special circumstances
with symmetries, the arguments of the functional $J(\cdot)$ may be
expressed as a function of a smaller group of variables to define
the landscape. For example, consider the state-to-state quantum
transition landscape $J[\varepsilon(\cdot)]=|\langle
f|U[\varepsilon(\cdot),t_f]|i\rangle|^2$, where
$U[\varepsilon(\cdot),t_f]$ is the system propagator at time $t_f$
under control $\varepsilon(\cdot)$ at going from the initial state
$|i\rangle$ to the target state $|f\rangle$. The landscape
function can be rewritten as $J(U)=|\langle f|U|i\rangle|^2$ over
the $N^2$-dimensional unitary transformation group manifold
$\U(N)$, or more compactly, $J(\psi)=|\langle f|\psi\rangle|^2$
over the unit sphere in the $N$-dimensional Hilbert space of
quantum states $|\psi\rangle$. A general connection between such
landscapes is shown by the following theorem:
\begin{theorem}\label{th}
Let $L(x)$ be a landscape defined on some differential manifold
$\X$. If $x$ is a function $x=h(y)$ of $y$ in some other
differential manifold $\Y$ with ${\rm dim~}\Y\geq {\rm dim~}\X$,
and $h$ is locally surjective on a neighborhood of some point
$y_0\in \Y$, i.e., the Jacobian has full rank:
\begin{equation}\label{surjective}
{\rm rank}\,\,\frac{\dd h}{\dd y}\Big |_{y=y_0}={\rm dim}\, \X\Big
|_{x=h(y_0)}
\end{equation}
in some local coordinate system. Then $y_0$ is critical for
$L\circ h$ in $\Y$ if and only if $x_0=h(y_0)$ is critical for $L$
in $\mathcal{X}$, and they have an identical number of positive
and negative Hessian eigenvalues at $y_0$ and $x_0$, respectively.
Moreover, if the inverse image $h^{-1}(x_0)$ of every critical
point $x_0$ is connected, then the connected components of their
critical manifolds are one-to-one between the two landscapes.
\end{theorem}
{\bf Proof:} From the chain rule \begin{equation}\label{chain}
\frac{\dd (L\circ h)}{\dd y}= \frac{\dd L}{\dd x}\cdot\frac{\dd
{h}}{\dd y},\end{equation}
 the rank condition (\ref{surjective})
guarantees that the gradient of $L\circ h$ vanishes at $y_0$ if
and only if the gradient of $L$ vanishes at $x_0$, hence $y_0$ is
critical if and only if $x_0$ is critical. Moreover, the Hessian
quadratic forms at the critical points are related by:
$$\frac{\dd^2 (L\circ h)}{\dd y^2}=\left(\frac{\dd h}{\dd y}\right)^\dagger\cdot \frac{\dd^2 L}{\dd  x^2}\cdot \frac{\dd h}{\dd y}.$$
The right hand side is a congruent transformation on the Hessian
quadratic form of $L$ at $x$ under the same full-rank condition
(\ref{surjective}), which preserves the signs of the nonzero
Hessian eigenvalues. Provided that the Hessian form is sufficient
to determine the types of critical points for both functions, then
any maximum (minimum, saddle) point in $\Y$ will be mapped to a
maximum (minimum, saddle) point in $\X$.

Every connected critical manifold of $L\circ h$ in $\Y$ must
produce a connected one of $L$ in $\X$ because the mapping $h:
\X\rightarrow \Y$ is continuous, so the number of critical
components in $\X$ must be no larger than those in $\Y$. In a
general setting, the number of connected components and their
simple (multiple) connectedness may not be preserved. Moreover,
let $y_1$ and $y_2$ belong to two connected critical submanifolds
in $\Y$, respectively, that are mapped into the same point $x_0$
in a connected critical component in $\X$. They must lie within
the connected component $h^{-1}(x_0)$, which is connected by
assumption. Therefore, the two critical submanifolds in $\Y$ can
be connected, and hence every connected component in $\X$ must be
mapped from a unique connected critical submanifold in $\Y$, i.e.,
the connected critical submanifolds in $\X$ and $\Y$ are
one-to-one. End of proof.

The rank condition is essential to the viability of lifting the
landscape analysis results in a kinematic picture (i.e., on the
unitary group irrespective to the system dynamics) to the control
landscape on the space of control fields, which has been adopted
in prior studies \cite{rabitz6,rabitz10,tak-san}. Here, the
full-rank property corresponds to the local controllability along
the trajectory driven by some reference control field. If the
reference control steers the system propagator to some (kinematic)
landscape critical point in $\U(N)$ and local controllability
holds, it must also be critical and possess the same optimality
status. Such controls are conventionally called regular extremal
controls \cite{BenMon2003}). According to (\ref{chain}), a
rank-deficient landscape lifting can lead to a critical control
field as well, which is called a singular extremal control that
may have no correspondence to any kinematic critical point in
$\U(N)$. So, the critical topology in the space of control fields
depends on both regular and singular extremal controls. The
optimality of singular extremals can generally be excluded
\cite{BenMon2003,WuRab2007}, and hence does not affect the search
of optimal controls. Therefore, it is sufficient to consider only
the normal extremal controls as the inverse image of the kinematic
critical points in $\U(N)$ via landscape lifting. This greatly
simplifies the landscape analysis by merely working in the
kinematic picture.

\section{Quantum Ensemble Landscapes of Observable Preparation}
Theorem \ref{th} provides a useful means to obtain the critical
submanifold of an unknown landscape by identifying it with the
image of some known landscape via a faithful lifting that
satisfies the conditions in Theorem 1. Consider the goal of
maximizing the expectation value of the observable $\theta$ of an
open quantum system with initial state $\rho$, which can be
formulated on the set of Kraus operators:
\begin{equation}\label{K-landscape}
    J(K)=\tr\left(\sum_{\alpha, \beta=1}^\lambda
K_{\alpha\beta} \rho K_{\alpha\beta}^\dagger
\theta\right)=\tr\{K(\rho\otimes I_\lmd) K^\dagger (\theta\otimes
I_\lmd)\}.
\end{equation}
We can lift this landscape via the mapping (\ref{map}) to the
following landscape over $\U(\lmd N)$:
\begin{equation}\label{U-landscape}
{J}(U)=\tr\left\{\F_\q(U)(\rho\otimes I_\lmd) \F^\dagger_\q(U)
\Theta\right\}=\tr\left(UPU^\dagger \Theta\right),
\end{equation}
where $P=\rho\otimes \q$ and $\Theta=\theta\otimes I_\lmd$.
Because $\K[\q,N]$ is homeomorphic to the homogeneous space of
$\U(\lmd N)$ as shown in (\ref{coset}), the properties in the
theorem above are naturally guaranteed from the facts that (a) the
mapping of a canonical projection is locally surjective
\cite{helgason} and (b) the inverse image of every Kraus operator
$K$ is homeomorphic to its coset in $\U(\lmd N)$ and hence is
connected. Thus, complete information can be extracted about the
critical landscape topology for open systems from the auxiliary
landscape (\ref{U-landscape}), which encompasses the composite
``system plus environment" model that targets the maximization of
the expectation value of the same system observable (minimization
may be treated just as well). We assume that the system is
controllable over the set of Kraus maps $\K[\q,N]$ generated from
an environmental initial state $\q$, while the controllability
over $\U(\lmd N)$ is not necessarily as demanding as it might seem
in practical physical circumstances since only a portion of the
composite system (e.g., just the system and its immediate local
environment) is likely required for manipulation. Furthermore,
most realistic laboratory control goals will admit as fully
acceptable just reaching some reasonable neighborhood of the
absolution maximum objective yield.

Note that the controllability criterion does not give any
information about the control landscape for open quantum systems,
which might be extremely complex in presence of environmental
interactions. However, an inherent property revealed about the
landscape for open systems can be immediately observed from the
above landscape lifting analysis. The open system landscape is
always devoid of {\it false traps} whenever the open system
dynamics is fully controllable, because the equivalent landscape
for the closed composite quantum system was proven to always
possess this basic novelty \cite{mike,rebing} owing to the
linearity and unitarity of quantum system dynamics. This result is
of fundamental importance for understanding the evident ease of
identifying effective control fields in the presence of
uncertainties
\cite{rabitz9,meshulach,levis,bixner,rabitz6,tak-san,mike2}
through the guarantee of not being caught by any false traps using
suitable search algorithms in the laboratory with flexible control
fields. In addition, the ability to decrease the entropy in the
system state shows that open system controlled dynamics can
outperform control of the corresponding isolated system without
introducing any false traps, as will be seen later.

The details of the landscape topology is affected by the initial
environmental state, which can be resolved via the equivalent
landscape (\ref{U-landscape}) as well. Other work \cite{rebing}
developed the technique of contingency tables to characterize
critical submanifolds. To be specific, suppose that $P$ has $r$
distinct eigenvalues with degeneracy degrees $m_1,\cdots,m_r$ and
$\Theta$ has $s$ distinct eigenvalues with degeneracy degrees
$n_1,\cdots,n_s$. Then every critical submanifold of
(\ref{U-landscape}) corresponds to a contingency table shown in
Table I.
\begin{center}
\begin{table}[h]
\begin{tabular}{c|ccc}
\hline
    & $~n_1~$ & $\cdots$ & $~n_s~$ \\
  \hline
  $~m_{1}~$ & $~c_{11}~$ & $\cdots$ & $~c_{1s~}$ \\
  $\vdots$ & $\vdots$ & $\ddots$ & $\vdots$ \\
  $m_{r}$ & $c_{r1}$ & $\cdots$ & $c_{rs}$ \\
  \hline
\end{tabular}\\
\tabcaption{The contingency table of the composite system. The
$c_{ij}$ entries satisfy the constraints $\sum_ic_{ij}=m_j$ and
$\sum_jc_{ij}=n_i$.}
\end{table}
\end{center}
Hence, the determination of critical submanifolds can be reduced
to a combinatorial problem of enumerating all possible contingency
tables. The dimension and the numbers of positive and negative
eigenvalues of the Hessian matrix can also be computed by these
contingency tables.

Considering the composite system plus an environment, the marginal
conditions can be determined by the degeneracy structures of
$\rho$, $\q$ and $\theta$. Suppose that $\rho$ has $r$ distinct
non-negative eigenvalues $1\geq p_1>\cdots
>p_{r-1}>p_r=0$ with multiplicities $d_1,\cdots,d_{r}$ ($d_r$ can be zero
corresponding to no zero eigenvalues in $\rho$); $\q$ has $m$
eigenvalues $1\geq q_1>\cdots>q_m=0$ with multiplicities
$f_1,\cdots,f_m$ ($f_m$ can be zero corresponding to no zero
eigenvalues in $\q$). The distinct eigenvalues of $\theta$ are
$r_1>\cdots>r_s$ with multiplicities $e_1,\cdots,e_s$. For
simplicity, we assume no accidental degeneracies in
$\rho\otimes\q$, i.e., $p_iq_j\neq p_kq_\ell$ for arbitrary $1\leq
i\neq k<r$ or $1\leq j\neq \ell<m$. In this case, the degeneracy
structure of the initial state $P=\rho\otimes \q$ is
$$D_{11}=d_1f_1,\,\,\cdots,D_{ij}=d_if_j\,\,\cdots,D_{r-1,m-1}=d_{r-1}f_{m-1},$$
corresponding to nonzero eigenvalues $p_iq_j$, where
$i=1,\cdots,r$ and $j=1,\cdots,m$, while the zero eigenvalue of
$P$ has multiplicity $D_{rm}=d_r\lmd+f_m N-d_{r}f_m$. The
observable $\Theta=\theta\otimes I_\lmd$ possesses the same group
of eigenvalues as $\theta$ whose degeneracy indices are multiplied
by $\lmd$, i.e.,
$$E_1=\lmd e_1,\,\,\cdots,\,\,E_{s-1}=\lmd e_{s-1},\,\,E_r=\lmd e_s.$$
The degeneracy structures define the contingency Table II, where
the row and column sums to be satisfied are
$(D_{11},\cdots,D_{r-1,m-1},D_{rm})$ and $(E_1,\cdots,E_s)$.
\begin{center}
\begin{table}[h]\label{open}
\begin{tabular}{c|cccc}
\hline
    & $~E_1~$ & $~E_2~$ & $\cdots$ & $~E_s~$ \\
  \hline
  $~D_{11}~$ & $~c_{111}~$ & $~c_{112}~$ & $\cdots$ & $~c_{11s~}$ \\
  $\vdots$ & $\vdots$ & $\vdots$ & $\ddots$ & $\vdots$ \\
  $~D_{r-1,m-1}~$ & $c_{r-1,m-1,1}$ & $c_{r-1,m-1,2}$ & $\cdots$ & $c_{r-1,m-1,s}$ \\
  $D_{rm}$ & $c_{rm1}$ & $c_{rm2}$ & $\cdots$ & $c_{rms}$ \\
  \hline
\end{tabular}\\
\tabcaption{The contingency table of the composite system.}
\end{table}
\end{center}


\subsection{The Full Control Landscape of Kraus Maps}

Consider the control landscape of all possible Kraus maps. As
discussed previously, this situation can be equivalently
associated with a special $N^2$-dimensional environment which is
initially in a pure state $\q$ that has only one nonzero
eigenvalue. The associated contingency Table III shows that $P$
has the same group of eigenvalues as $\rho$, except for the
increase in the number of zero eigenvalues by $N^3-N$.

\begin{center}
\begin{tabular}{c|ccc}
\hline
    & $N^2 e_1$ &  $\cdots$ & $N^2 e_s$ \\
  \hline
  $d_1$ & $c_{11}$ & $\cdots$ & $c_{1s}$ \\
  $\vdots$ & $\vdots$ &  $\ddots$ & $\vdots$ \\
  $d_{r-1}$ & $c_{r-1,1}$ &  $\cdots$ & $c_{r-1,s}$ \\
  $d_r+(N^3-N)$ & $c_{r1}$ & $\cdots$ & $c_{rs}$ \\
  \hline
\end{tabular}\\
\tabcaption{The contingency table for the landscape with pure
environmental initial states. }
\end{center}

Because of the marginal conditions, the first $r-1$ rows uniquely
determine the contingency table provided that the sum of the first
$r-1$ elements in each column does not exceed the corresponding
column sum. This criterion is automatically satisfied in Table III
because
$$c_{1j}+\cdots+c_{r-1,j}\leq d_1+\cdots+d_{r-1}\leq N<N^2e_j,\quad j=1,\cdots,s.$$
Thus, the column-sum constraints are actually ineffective, which
implies that the number $\mathcal{N}$ of contingency tables
depends only on the number of ordered partitions of $d_1,\cdots,
d_{r-1}$, i.e.,
\begin{equation}\label{nmax} \mathcal{N}=\prod_{i=1}^{r-1}
\frac{(d_i+s-1)!}{d_i!(s-1)!}.
\end{equation}
The dependence of $\mathcal{N}$ on the degeneracy indices of the
nonzero eigenvalues of $\rho$ manifests itself in the contribution
of the entropy of the system to the landscape topology.
Apparently, the diverse nature of the initial state distributions
tends to give rise to more critical submanifolds for large $N$.
The observable $\theta$ affects the number of critical
submanifolds only by its number of distinct eigenvalues. A
physical interpretation of this number is the resolution required
to measure the target observable, whose low value facilitates the
control and measurement, and hence reduces the complexity of the
control landscape despite the environment being present.

Concerning the local topological landscape properties, a prime
interest is the number of nonzero eigenvalues at the global
optima, which represents the number of effective search directions
close in that neighborhood. The associated contingency Table IV
has nonzero entries only in its first column and last row.
Applying the formulas developed in \cite{rebing}, we can calculate
the number of nonzero Hessian eigenvalues (all negative) as
$$\M_-=2N^2(N-d_r)(N-e_1).$$ This characteristic is proportional to the number of zero eigenvalues
in $\rho$ and the number of eigenvalues smaller than $e_1$ in the
observable $\theta$. Again, the low entropy of $\rho$ helps to
reduce the number of search directions towards the global optima,
and improves the robustness of this critical manifold. Similarly,
the large size $e_1$ of the eigenspace of the maximal eigenvalue
of $\theta$ also facilities the search near the optima.

\begin{center}
\begin{tabular}{c|ccc}
\hline
    & $N^2 e_1$ &  $\cdots$ & $N^2 e_s$ \\
  \hline
  $d_1$ & $d_1$ & $\cdots$ & $0$ \\
  $\vdots$ & $\vdots$ &  $\ddots$ & $\vdots$ \\
  $d_{r-1}$ & $d_{r-1}$ &  $\cdots$ & $0$ \\
  $d_r+(N^3-N)$ & $N^2e_1-N+d_r$ & $\cdots$ & $N^2e_s$ \\
  \hline
\end{tabular}\\
\tabcaption{The contingency table for the global extremum of the
open system control landscape.}
\end{center}

\subsection{Control Landscapes with a Thermal Environment}

This section considers the control landscapes under the common
situation that the environment is initially in a thermal
equilibrium state
\begin{equation}\label{thermal state}
    \q(T)=\frac{\exp(-H_E/kT)}{\tr [\exp(-{H_E}/{kT})]},
\end{equation} where $T$ is the temperature and $k$ is the
Boltzmann constant. For simplicity, we suppose that the
environmental Hamiltonian has a non-degenerate spectrum, i.e.,
$f_1=\cdots=f_m=1$. Although under most circumstances the
associated Kraus representations do not cover all possible maps,
the landscape still exhibits no false traps according to the
analysis. Appendix A estimates the approximate numbers of critical
submanifolds under different circumstances. In the limit of zero
temperature, the number reaches its maximum value as shown in
equation (\ref{nmax}), and remains constant when $\lmd$ is
sufficiently large. As the temperature increases, more critical
submanifolds grow out from the relatively flat low-temperature
landscape. The set of critical submanifolds are identical for all
finite temperature values and their number increases exponentially
with the size of the environment, while the geometry (e.g., the
curvature near the critical submanifolds) varies with the
temperature. In the limit of infinite temperature, the landscape
critical submanifolds merge into a smaller group whose number
increases only polynomially with the size of the environment.

Open system control can also extend the scope of manipulation over
the expectation value of the target observable, which is reflected
by the dynamical range of the control landscape defined as $\Delta
J= J_{\max}-J_{\min}$. For closed systems that evolve from pure
states, the attainable landscape values range from the maximal
eigenvalue $r_1$ of the target observable $\theta$ to its minimal
eigenvalue $r_s$, i.e., $\Delta J= r_1-r_s$; the range is usually
further restricted in the case of mixed states and in the presence
of external disturbances \cite{bounds}. The dynamical range of
open quantum systems may be enhanced through control over the
environment and its interactions with the system. Fig.1 examines
the dynamical range of a two-level quantum system, where the
observable is chosen as the Pauli operator $\sigma_z$ whose
eigenvalues are $r_1=1/2$ and $r_2=-1/2$. The landscape always has
the ideal dynamical range $\Delta J_{\max}=1$ at zero temperature,
and shrinks when the temperature increases. Comparing the trends
for coupling to different environments, we see that the dynamical
range is wider when the open dynamics is fully controllable and
when interacting with larger environments. In the limit of
infinite temperature, the dynamical range approaches the lower
bound for that of the isolated system. In that case the enhanced
controllability of the open dynamics benefits the control
landscape very little. The distinction between the two groups of
curves that differ by the initial system state $\rho$ shows that
lower-entropy helps to extend the dynamical range, i.e., it
facilitates higher-quality controls. The degree of attainable
control over the environment will be an application specific
matter. Importantly, it is reasonable to expect that only the
neighboring environment to the system will likely need to be
controlled, and this prospect may open up practical conditions for
effective use of the environment to aid in the control of system
dynamics.
\begin{figure}[h]
\centerline{
\includegraphics[width=5in,height=3.5in]{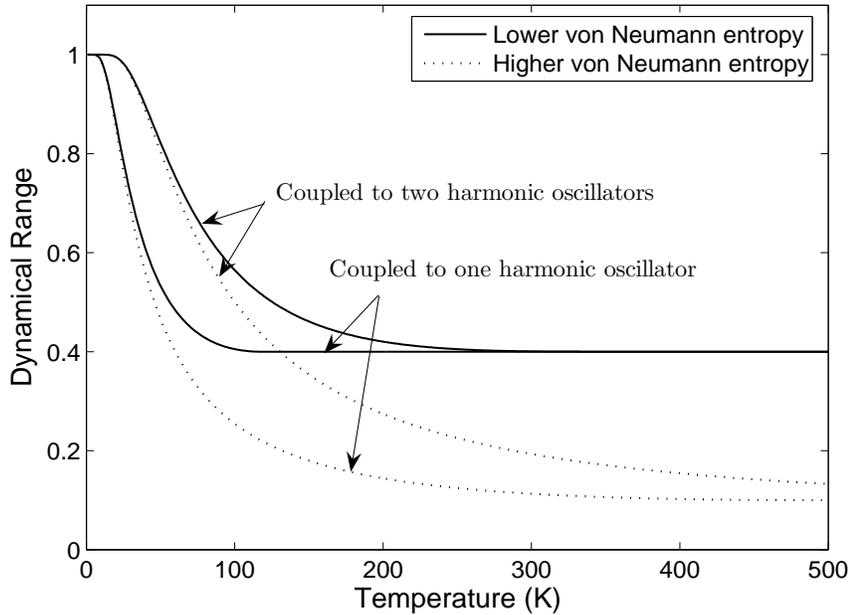}
} \caption{ Dynamical range of a two-level system in different
thermal environments including one and two harmonic oscillators.
The eigenvalues of $\theta$ are $r_1=1/2$ and $r_2=-1/2$. The
populations of $\rho$ with higher entropy are $p_1=0.55$ and
$p_2=0.45$, while $p_1=0.70$ and $p_2=0.30$ for $\rho$ with lower
entropy. }\label{thermal}
\end{figure}

We illustrate the behavior of the search for optimal dynamical
control of the above two-level spin-$\frac{1}{2}$ system
interacting with some relatively small environments represented by
a spin-$\frac{\lmd-1}{2}$ system (here we choose $\lmd=6$). The
total system Hamiltonian is $H_{\rm
total}=H_0+\varepsilon(t)\sigma_x\otimes I_\lmd$ with
$$H_0=\omega_0\sigma_z\otimes I_\lmd+I_2\otimes
\omega_eJ_z+\gamma\sum_{\alpha=x,y,z}(\sigma_\alpha\otimes
J_\alpha ),$$ where the $\sigma$'s are the standard Pauli matrices
and the spin operators $J_x,J_y,J_z$ form a basis of the $SO(3)$
Lie algebra and possess a $\lmd$-dimensional irreducible unitary
representation of the $SO(3)$ Lie algebra basis. The environment
is initially in a thermal state (\ref{thermal state}), where
$H_e=\omega_eJ_z$.

Although the control field only acts directly on the system
itself, it can be demonstrated that the rank of the Lie algebra
spanned by $H_0$ and $\sigma_x\otimes I_\lmd$ equals that of the
Lie algebra of $\U(2\lmd)$, which implies that the composite
system is controllable via the system-environment coupling given a
sufficiently large final time \cite{rabitz3}. Fig.2 shows the
maximization of the cost function (\ref{K-landscape}) using the
conjugate gradient algorithm, and the resulting system entropy at
each iteration step. As already shown in Fig.1, the dynamical
search finds fields that transfer population to the target level
beyond the capability of the closed system control landscape at
different temperatures. Moreover, the optimal control fields also
purify the system as shown in the decrease of the von Neumann
entropy in Fig.2. Hence, a clever use of the environment may
improve the accessible dynamical range on the system landscape.
Fig.3 shows the resulting optimal control field with and without
the presence of the environment (i.e., for the case of $T=300K$).
With a relatively small environment, the resonant mode for the
state transition arising from the system free Hamiltonian is still
dominant in the optimal control field. The entropy in the system
state drops slightly more quickly when the temperature of the
environment is higher before optimization, i.e., the controlled
system becomes cooler when initially bathed in a hotter
environment. Although such behavior are physically allowed, the
phenomenon is still counter-intuitive. This may be an algorithm
and objective-dependent issue with the present goal maximization
of the expectation value of $\theta$ without any specific
consideration of minimizing the system entropy.

\begin{figure}[h]
\centerline{
\includegraphics[width=3in,height=2.5in]{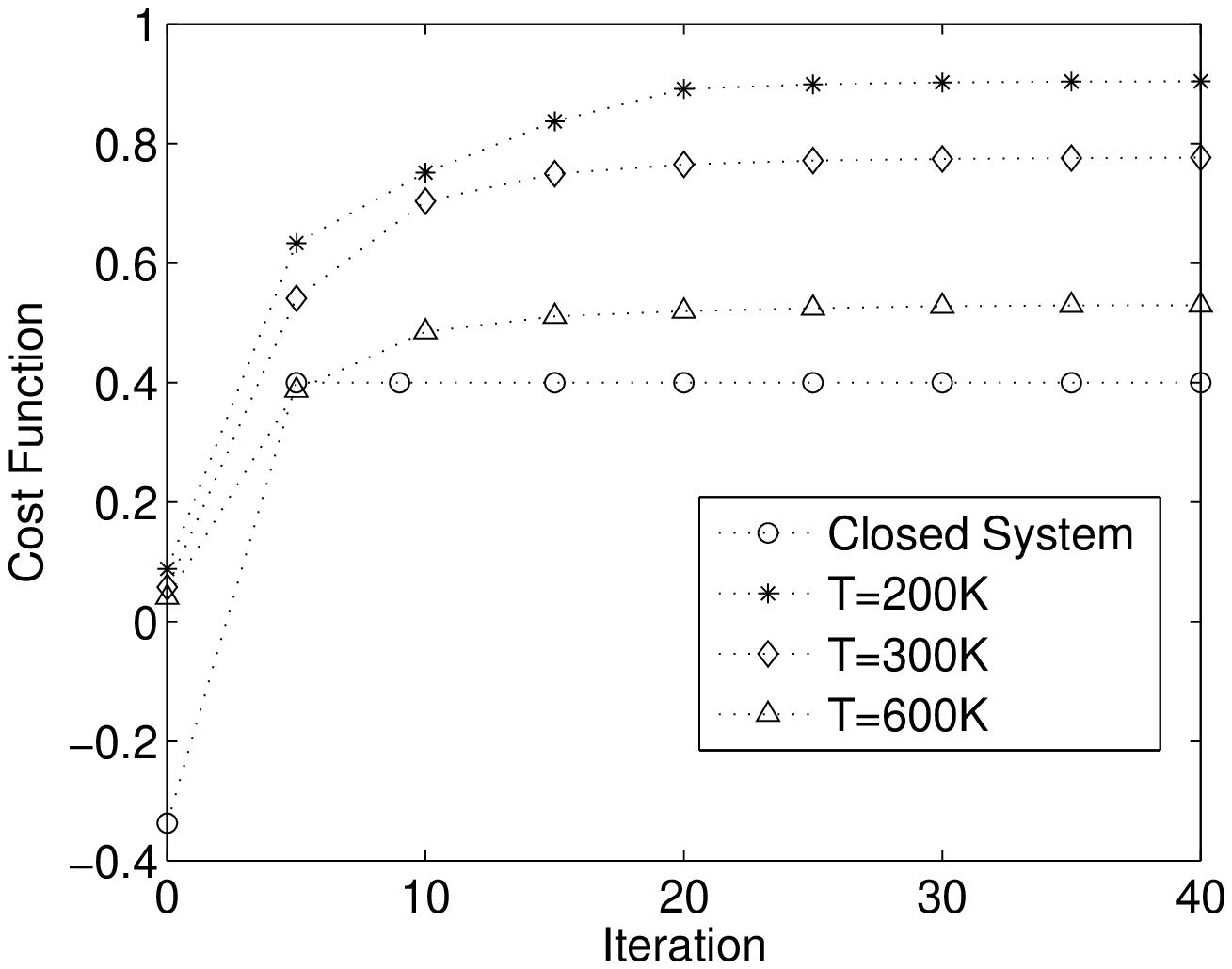}
\includegraphics[width=3in,height=2.5in]{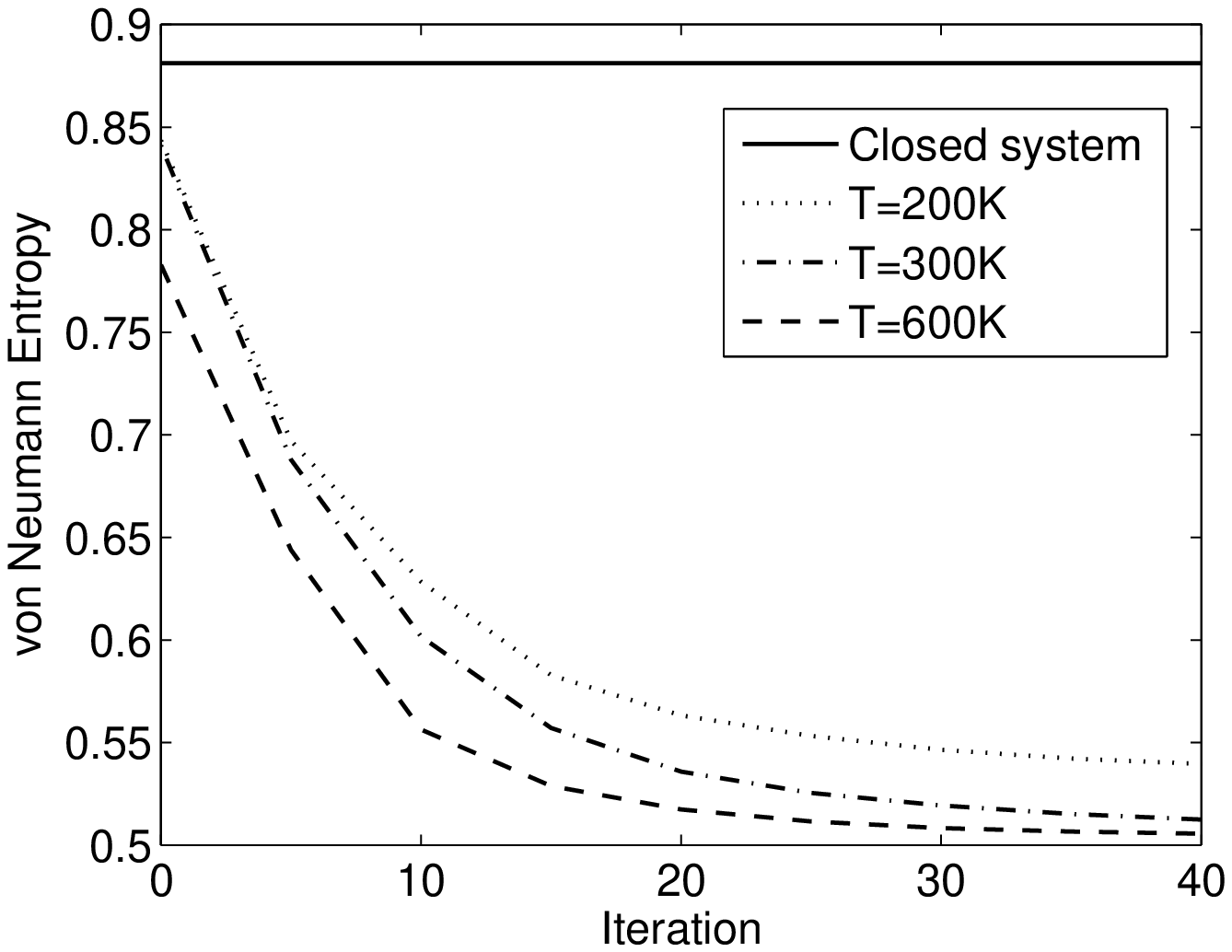}
} \caption{ Dynamical search for optimal controls in a two-level
open quantum system, where the dimension of the associated
environment is $\lmd=6$. The observable is
$\theta=\sigma_z$.}\label{thermaldy2}
\end{figure}

\begin{figure}[h]
\centerline{
\includegraphics[width=6in,height=5in]{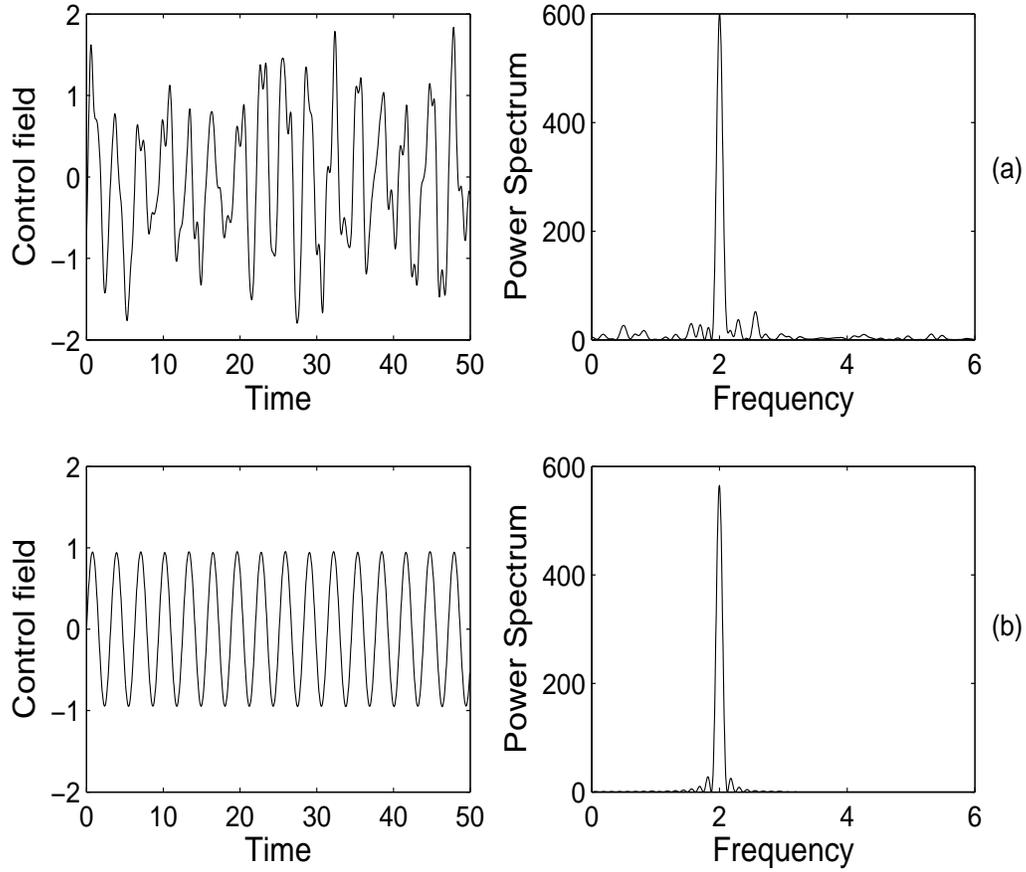}
} \caption{ Control fields for open quantum control systems: (a)
in presence of the environment initially at temperature $T=300K$;
(b) in absence of the environment. }\label{thermaldy2}
\end{figure}

\section{conclusion}
This work presents an investigation of control landscape topology
for open quantum systems. The analysis was performed with the aid
of landscape lifting and reveals several inherent properties of
the open system control landscapes. The findings here provide a
basis to expect that (a) the search for an optimal control in the
laboratory may not be significantly hindered by the presence of an
environment, (b) no traps exist to limit the control yield, and
(c) in favorable cases, the environment can aid the control
outcome. These conclusions are consistent with the observed
relative ease of attaining control over quantum systems in the
laboratory, including those of high complexity in condensed phase
environments.

A lack of full controllability may turn some of the saddle
critical points for controllable systems into local maxima, and
thus generate false traps. Moreover, the influence of singular
extremal controls may also be significant when the system is not
controllable, because they often appear at the boundary of the
reachable set and can become optimal solutions to the control
problem. Further studies should take into account such limiting
factors as well as (1) uncontrollable degrees of freedom and (2)
the employment of restricted control fields, which might generate
apparent false traps in the search for effective controls.

\begin{acknowledgments}
The authors acknowledge support from the DOE.
\end{acknowledgments}

\appendix
\section{Estimation of the number of critical submanifolds}
An estimate can be given for the number of critical submanifolds,
whose precise enumeration is generally NP-hard computable
\cite{bender,boon}. The following formula matches well to the
actual number of critical submanifolds when there are many rows of
the contingency table and some of the row sums are large. Suppose
that $r\geq s$, i.e., $\theta$ has higher degeneracies, then we
have the Gaussian formula \cite{gail} for the number of critical
submanifolds of the isolated system:
\begin{equation}\label{dj}
A_0\approx\frac{s^{\frac{1}{2}}}{{(2\pi\beta})^{\frac{s-1}{2}}}M\exp\left(-\frac{Q}{2\beta}\right),
\end{equation}
where
$$Q=\sum_{j=1}^sE_j^2-\frac{N^2}{s},\quad M=\prod_{i,j}{\frac{(D_{ij}+s-1)!}{D_{ij}!(s-1)!}},
\quad \beta=\sum_{i=1} \frac{D_{ij}(D_{ij}+s)}{s(s+1)}.$$

Consider the following circumstances:

(1) For systems with thermal environments at finite nonzero
temperature, the marginal sums of the contingency tables are
$$D_{ij}=d_i,\,\, E_k=\lmd e_k, \quad i=1,\cdots,r;j=1,\cdots,m;k=1,\cdots,s,$$from which we
obtain the approximate number of critical submanifolds associated
with the $\lmd$-dimensional thermal environment:
\begin{equation}\label{dj}
A_\lmd\approx\frac{s^{\frac{1}{2}}M_0^\lmd}{{(2\pi\lmd\beta})^{\frac{s-1}{2}}}\exp\left(-\frac{\lmd
Q}{2\beta}\right),\quad
M_0=\prod_{i}^{r-1}{\frac{(d_i+s-1)!}{d_i!(s-1)!}}.
\end{equation}
For large environment dimension $\lmd$, the number of landscape
saddles increases exponentially with $\lmd$.

(2) In the limit of zero temperature, the environmental
equilibrium state reduces to a pure state. In this case, the
number of critical submanifolds remains constant when $\lmd$ is
sufficiently large, and the precise number is given by the
equation (\ref{nmax}).

(3) In the limit of infinite temperature,
$\q_\infty=\lmd^{-1}I_\lmd$ becomes a completely mixed state,
which leads to the marginal sums of the contingency tables:
$$D_{i}=\lmd d_i,\,\, E_j=\lmd e_j, \quad i=1,\cdots,r;j=1,\cdots,s.$$One can show that for large $\lmd$,
$$Q_\lmd=\lmd^2 Q,\quad M_\lmd\approx \prod_{i=1}^r{\frac{(\lmd d_i)^{s-1}}{(s-1)!}},\quad \beta_\lmd\approx \lmd^2 \beta,$$
which leads to
\begin{equation}\label{dji}
A_\lmd\approx
\left[\frac{s^{\frac{1}{2}}e^{-\frac{Q}{2\beta}}}{{(2\pi\beta})^{\frac{s-1}{2}}}\prod_{i=1}^r{\frac{d_i^{s-1}}{(s-1)!}}
\right]\lmd^{(r-1)(s-1)}.
\end{equation}
Therefore, the approximate number of critical submanifolds
increases polynomially with $\lmd$ in the limit of infinite
temperature.

\newpage


\end{document}